\documentclass[conference]{IEEEtran}
\IEEEoverridecommandlockouts

\usepackage[utf8]{inputenc}
\usepackage[T1]{fontenc}

\usepackage{amsmath,amssymb}
\usepackage{booktabs}
\usepackage{tabularx}

\usepackage{graphicx}
\usepackage{svg}
\usepackage{cite}
\usepackage{float} 
\usepackage{alltt}
\usepackage{balance}
\usepackage{enumitem}

\usepackage{fancyhdr} 

\usepackage[hidelinks]{hyperref}

\makeatletter
\def\ps@IEEEtitlepagestyle{%
  \def\@oddhead{}%
  \def\@evenhead{}%
  \def\@oddfoot{\parbox{\textwidth}{\scriptsize \copyright 2026 IEEE. Personal use of this material is permitted. Permission from IEEE must be obtained for all other uses, in any current or future media, including reprinting/republishing this material for advertising or promotional purposes, creating new collective works, for resale or redistribution to servers or lists, or reuse of any copyrighted component of this work in other works. Cite this article as follows: N. H. Fahim, A. U. Muhib, M. S. Rahman, ``Protein-Based Fish Species Identification: Dataset, Models, and Insights from Native Bangladeshi Fish,'' \textit{2026 IEEE 2nd International Conference on Quantum Photonics, Artificial Intelligence, and Networking (QPAIN)}, Chittagong, Bangladesh, 2026. DOI: \href{https://doi.org/10.1109/QPAIN69676.2026.11546620}{10.1109/QPAIN69676.2026.11546620.}}}%
  \def\@evenfoot{}%
}
\makeatother

\title{Protein-Based Fish Species Identification: Dataset, Models, and Insights from Native Bangladeshi Fish}

\author{
    \IEEEauthorblockN{Md Nasiat Hasan Fahim}
    \IEEEauthorblockA{Shahjalal University of Science \\ and Technology, \\
    Sylhet \\
    Email: \href{mailto:nhfahim18@gmail.com}{nhfahim18@gmail.com}}
    \and
    \IEEEauthorblockN{Md. Abid Ullah Muhib}
    \IEEEauthorblockA{Shahjalal University of Science \\ and Technology, \\
    Sylhet \\
    Email: \href{mailto:uusshas12@gmail.com}{uusshas12@gmail.com}}
    \and
    \IEEEauthorblockN{Mohammad Shahidur Rahman\textsuperscript{*}}
    \IEEEauthorblockA{Shahjalal University of Science \\ and Technology, \\
    Sylhet \\
    Email: \href{mailto:rahmanms@sust.edu}{rahmanms@sust.edu}}
}

\begin{document}
\maketitle

\begin{abstract}
Correct identification of fish species is highly significant for food security, economic development, and climate resilience in Bangladesh. Protein sequences directly reflect functional and evolutionary constraints which are important for species authentication and biodiversity monitoring. Yet there exists no benchmark for native Bangladeshi fish species identification from protein sequence. In this study, we addressed this gap by introducing the first curated dataset for nine native Bangladeshi fish species of 2845 high quality protein sequences. We also established the first protein sequence classification baseline for this domain through a systematic benchmarking of seven architectural paradigms. Moreover, we propose a realistic deployable novel hybrid architecture of MoftiCNN and Transformer with Terminal-Aware Positional-Encoding (MotifCNN-Transformer+TA-PE). Our novel architecture achieves 79.80\% accuracy with macro-$F_1$ of 0.80. The highest 83.04\% accuracy is achieved by finetuned protein language model ProtBERT that has 420M parameters and requires dual 16GB GPUs for inference. According to McNemar's test, ProtBERT's 3.24\% accuracy gain over our MotifCNN-Transformer+TA-PE is statistically insignificant($p=0.1120$). Our novel architecture beats it among six of the nine classes in per class identification. Also our MotifCNN-Transformer+TA-PE is approximately 5$\times$ faster, 42$\times$ smaller, and supports 16$\times$ larger batch size than ProtBERT and has GPU free inference, making it more practical for deployment in resources constrained areas such as rural Bangladesh. Beyond this, our foundational work shows effects of phylogenetic relationships on sequence similarity and establishes pathways for fisheries management, food authentication and biodiversity conservation in South Asia's protein dependent economy. 
\end{abstract}


\section{Introduction}

Fish and fisheries are intricately connected to the socio-economic culture of Bangladesh. This country is the third largest country in inland fish production and is surrounded by 0.84 million hectares of closed inland waters, 3.86 million hectares of open inland waters, and a 710 km coastal belt. Fisheries sector contributes 3.50\% to GDP, and it's role of providing livelihoods to around 12\% population have made it a cornerstone of national food security and economic prosperity~\cite{r30}. Therefore, to support fish population and biodiversity management, drive economic sustainability and growth, combat commercial food fraud, safeguard human diet and well-being fish species identification is crucial.

 Several methods exist for fish identification, but they exhibit significant limitations. Neither DNA barcoding nor morphological methods can identify processed fish(e.g., dried, canned) where the primary DNA structure is disrupted and visual features are missing or degraded~\cite{ZHAO2024}. On the other hand, protein based methods can effectively authenticate processed fish~\cite{Meledina2025}. Not only that, protein based methods are also practically applicable for the identification of fish species~\cite{CHIEN2022}.

Inspite of this fundamental importance of protein based species identification methods, no curated dataset or computational benchmark on protein sequence exists for native Bangladeshi fish species. Addressing the gap, this study establishes the foundation for protein based fish species identification. Our key contributions are as follows.

\begin{itemize}
    \item The curation of a novel dataset of 2845 high quality protein sequences across nine economically and ecologically important native Bangladeshi fish species. The species are taken from both freshwater and saltwater.
    \item The development of first systematic benchmark by comparing seven architectural paradigms under unified experimental protocols ranging from classical machine learning methods to state of the art deep learning approaches.
    \item The introduction of a novel hybrid MotifCNN-Transformer architecture with Terminal-Aware Positional-Encoding achieving near state of the art accuracy while remaining 42$\times$ smaller, approximately 5$\times$ faster and supporting 16$\times$ larger batch size than the best performing ProtBERT with GPU free inference.
    \item The assessment of the effects of phylogenetic relationships on sequence similarity and a practical solution of deployable model in resource constrained regions.
\end{itemize}

\section{Related Work}

Traditional fish species identification relies on morphological characteristics and DNA barcoding techniques. Morphological methods analyze external features such as fin shape and body morphology~\cite{Irschick2017}. Morphology-based identification remains foundational in fisheries management, but suffer from poor detection accuracy, limited scalability, and failure to detect certain species~\cite{Adhikary2025}. DNA barcoding emerged as a molecular alternative using cytochrome c oxidase I (COI) sequences and requires sophisticated wet-lab infrastructure~\cite{Xing2018}. Recent studies show that DNA barcodes are ineffective for species identification~\cite{Quek2024}.

Early computational approaches used traditional machine learning with handcrafted features(e.g., amino acid composition) for protein based species identification~\cite{chakraborty2025}. They demonstrated proof of concept for protein based authentication but their limitation of capturing complex sequence patterns limited their performance~\cite{Ahmed2024}.

The advent of deep learning models has replaced this automated hierarchical feature extraction that captures complex patterns in protein sequences~\cite{Hein2025}. Recent work demonstrates that transformer-based models outperform traditional CNN/LSTM baselines~\cite{Bari2025}. Following this, protein language models (PLMs) like ProtBERT and ESM-2 achieve state of the art results but require substantial computational resources~\cite{Lin2023}.

Effective alternatives to these are hybrid architectures that combine convolutional feature extraction with attention mechanisms~\cite{Mahala2025}. MotifCNN can capture functional domains while maintaining computational traceability~\cite{sankar2019}. And transformer architectures can focus on evolutionarily conserved regions by leveraging self-attention mechanisms that give them the ability to discriminate closely related species~\cite{Choi2023}.

Existing work in Bangladesh focuses primarily on taxonomic checklists \cite{r29}, limited DNA barcoding studies \cite{r27}, or image-based identification \cite{r28}. Despite these progresses, there exists no curated protein sequence dataset or systematic benchmark for native Bangladeshi fish species. Current computational resources focus exclusively on global commercial species or rely on DNA based methods. Resource constrained deployment requirements are ignored by state of the art protein language models(pLMS). Finally, existing approaches do not account for phylogenetic relationships that fundamentally constrain sequence based discrimination among closely related species. 

\section{Dataset And Preprocessing}
Our dataset contains protein sequences of nine native Bangladehsi freshwater and saltwater fish species. These species were selected based on their cultural and economical importance. Rohi and mrigal are common all over the country, ilish is nationally significant, chital and sing are unique freshwater varieties, and rupchanda and sapla pata are from the Bay of Bengal. Protein sequences were directly collected from UniProtKB~\cite{UniProt2024} the Universal Protein Knowledgebase and NCBI~\cite{Sayers2025} the National Center for Biotechnology Information that formed a standardized and curated dataset for the identification of fish species. 

\begin{figure}[ht]
    \centering
    \includegraphics[width=\linewidth]{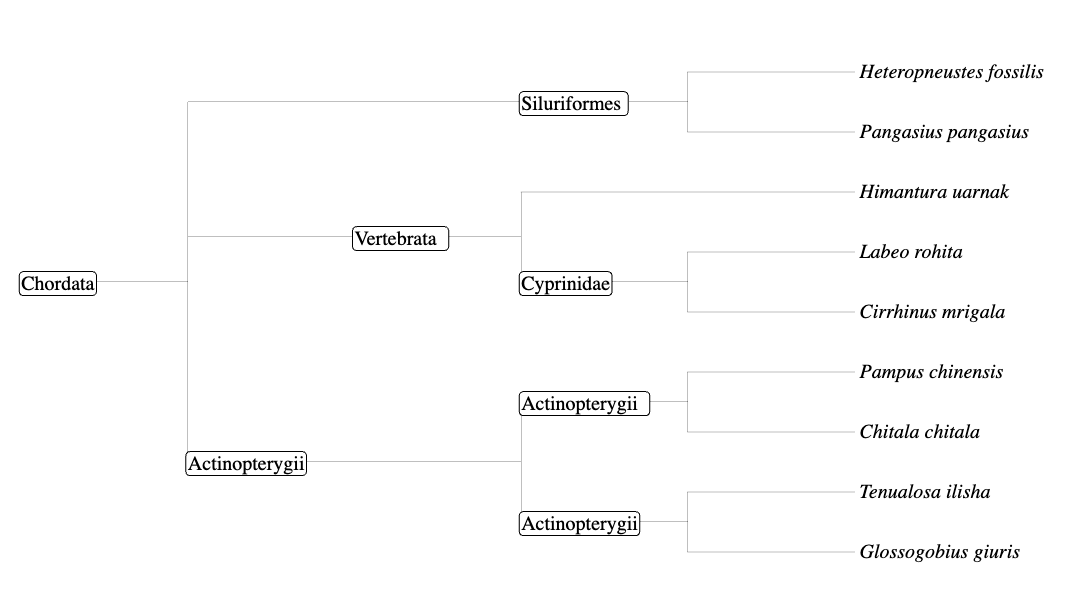}
    \caption{Near and distant phylogenetic relationships among the nine species, illustrating evolutionary constraints that impact sequence-based discriminability.}
    \label{fig:taxonomy}
\end{figure}

The final dataset consists of 2845 high quality protein sequences after rigorous quality control and preprocessing from nine fish species. These species share both near and distant phylogenetic relationships with a varying range of sequence similarity. Fig.~\ref{fig:taxonomy} represents these relationships of taxonomic lineages spanning from phylum to species.

\begin{figure}[ht]
    \centering
    \includegraphics[width=\linewidth]{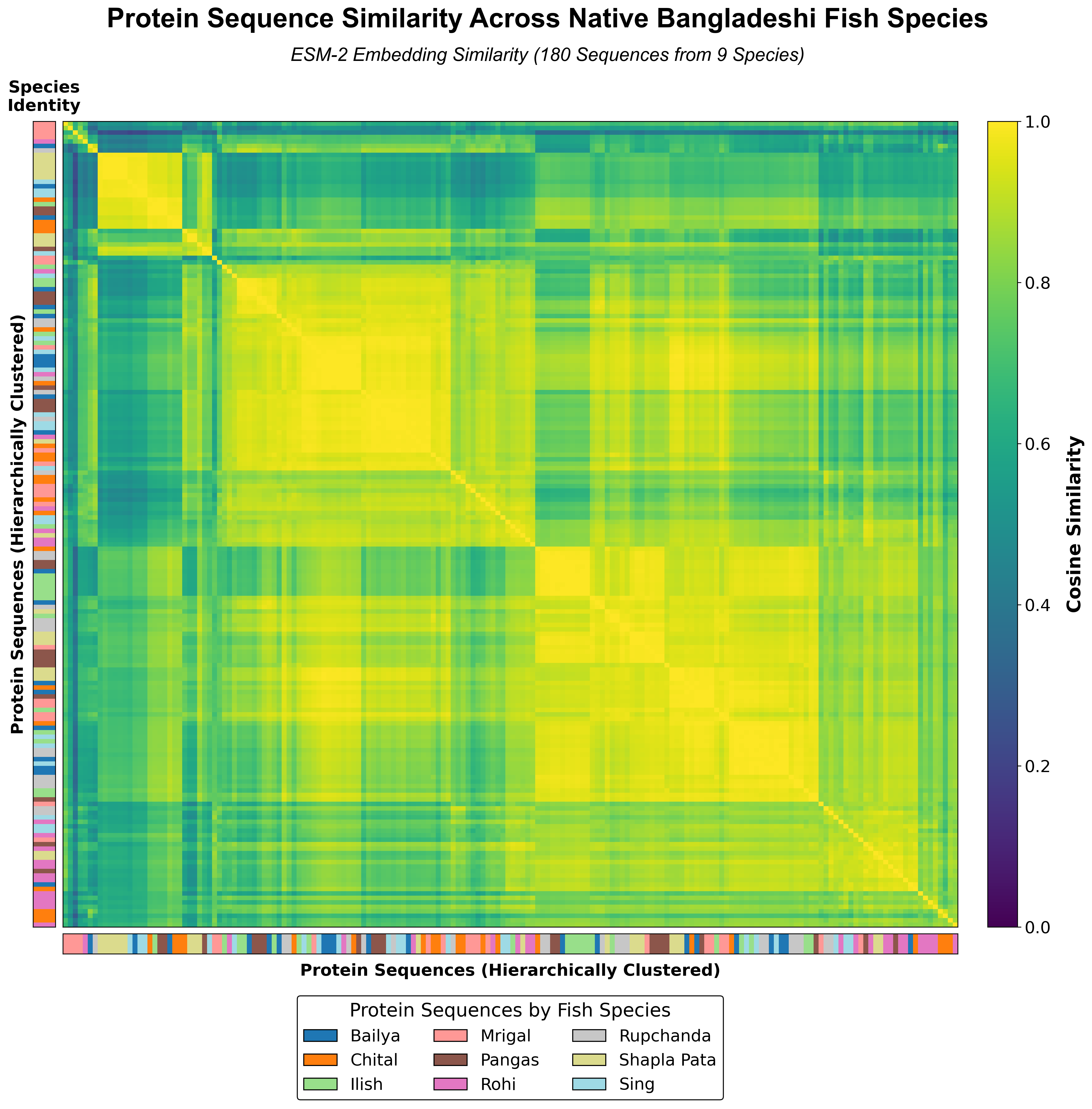}
    \caption{Protein sequence similarity heatmap for nine fish species containing 20 sequences from each species through random sampling using ESM-2 embeddings.}
    \label{fig:similarity}
\end{figure}

Fig.~\ref{fig:similarity} illustrates the high similarity of protein sequences among the nine species of the dataset. This similarity was calculated using ESM-2 protein language model embeddings through the random sampling of 20 sequences from each species. The heatmap shows that most of the sequences are nearly 80\% to 90\% similar. This makes species discrimination very difficult.

Table~\ref{tab:class_counts} shows the common names and scientific names of the nine species and their train/validation/test split of 70:15:15(2,016 training samples, 415 validation samples, and 414 test samples) with the count of the total number of sequences in the final dataset. This split is done using stratified splitting protocol to maintain class distribution. The table also shows the inherent class imbalance of the dataset ranging from 223 to 515 sequences.

\begin{table}[ht]
\centering
\setlength{\abovecaptionskip}{3pt}
\setlength{\belowcaptionskip}{2pt}
\caption{Distribution of sequences across species in our dataset, showing both common names used in Bangladesh and scientific nomenclature.}
\label{tab:class_counts}
\begin{tabular}{llrrrr}
\toprule
Common & Scientific name & Train & Val & Test & Total \\
\midrule
chital & Chitala chitala & 333 & 67 & 67 & 467 \\
mrigal & Cirrhinus mrigala & 365 & 75 & 75 & 515 \\
pangas & Pangasius pangasius & 215 & 46 & 45 & 306 \\
rupchanda & Pampus chinensis & 208 & 43 & 43 & 294 \\
ilish & Tenualosa ilisha & 185 & 38 & 37 & 260 \\
rohi & Labeo rohita & 209 & 44 & 44 & 297 \\
baliya & Glossogobius giuris & 180 & 37 & 36 & 253 \\
shapla pata & Himantura uarnak & 163 & 33 & 34 & 230 \\
sing & Heteropneustes fossilis & 158 & 32 & 33 & 223 \\
\bottomrule
\end{tabular}
\end{table}

To ensure data quality and biological validity our dataset curation process involved multiple stages: 1) Removal of exact duplicate sequences, 2) Exclusion of sequences by a length filter of 100AA(amino acids) to 800AA, 3) Removal of fragmented (completeness > 90\%) and missing (threshold = 5\%) sequences~\cite{UniProt2024}, 4) Removal of contaminated sequences with a threshold of 95\%~\cite{UniProt2024}, 3) Quantification of low complexity region with a tolerance rate of 20\%~\cite{MARSHALL2023}, 5) Exclusion of more than 2\% ambiguous sequences and remaining ambiguity mapping to 'X'~\cite{Berrow2021}, 6) Validation of saltwater sequences by calculating their isoelectric point (pI) score($\sim$4.2-4.8) and Aspartic + Glutamic acid(20\%) percentages~\cite{Paul2008,Oren2013}.

Our analysis reveals that length distribution of the sequences significantly heterogeneous ranging from 14AA to 2856AA with a right skewed distribution(skewness: 3.761). The sample mean of the sequences is 305.70$\pm$243.33. Proteins with less than 100AA are often fragmented and more than 800AA increases computational costs. The sample mean, fragmentation and computational costs justifies the length filter. The filter retains 96.10\% sequences and 93.76\% of total residues. The five most frequent residues (L: 13.33\%, A: 7.97\%, S: 7.03\%, I: 6.71\%, G: 6.60\%) exhibit varying distributions across species, suggesting that compositional features alone may provide discriminative power for species identification.

\section{Methodology}

We implemented and evaluate seven architectural paradigms spanning from traditional machine learning to state of the art protein language models(pLMs) under unified experimental protocols. 
Each model processes protein sequences ranging from 100AA to 800AA length that establishes performance boundaries and identifies optimal architectures for species discrimination.

1) N-gram + LightGBM: Our classical baseline extracts character n-grams (n=1–4) transformed via TF-IDF (max 5,000 features) and classified using LightGBM. This alignment free approach establishes a performance floor for comparison.

2) MotifCNN: Inspired by DeepLoc~\cite{DeepLoc2017}, this architecture employs multiscale 1D convolutions(kernel sizes: 1, 3, 5, 9, 15, 21, filters: 128 for each kernel) with embedding dimension of 128. Concatenated features undergo global max pooling followed by dropout(0.3) for classification. This approach uses local feature detection to make prediction.

3) CNN + BiLSTM: Following the full DeepLoc architecture~\cite{DeepLoc2017}, this model combines the same multi scale CNN with two bidirectional LSTM layers(128 hidden units each) and dropout(0.3). This configuration combines local feature extraction with explicit sequential modeling for species discriminates.

4) Standard Transformer: Our attention based baseline implements a 6 layer Transformer encoder(embedding dimension=256, 8 attention heads, feedforward dimension=1024, dropout=0.1) with sinusoidal positional encoding. This architecture uses attention mechanisms to capture global dependencies without sequential processing constraints.

5) Transformer with Terminal-Aware Positional-Encoding: This architecture is build upon the standard Transformer with identical hyperparameters and applies biologically motivated scaling to positional encoding that emphasize terminal regions. Scaling factors transition from $2.0\rightarrow 1.0$ across the first 50 residues(N terminus) and from $1.0\rightarrow 2.0$ across the final 50 residues(C terminus). Predictions of this model reflects the biological importance of terminal regions in protein function.

\begin{figure}[ht]
    \centering
    \includegraphics[width=\linewidth]{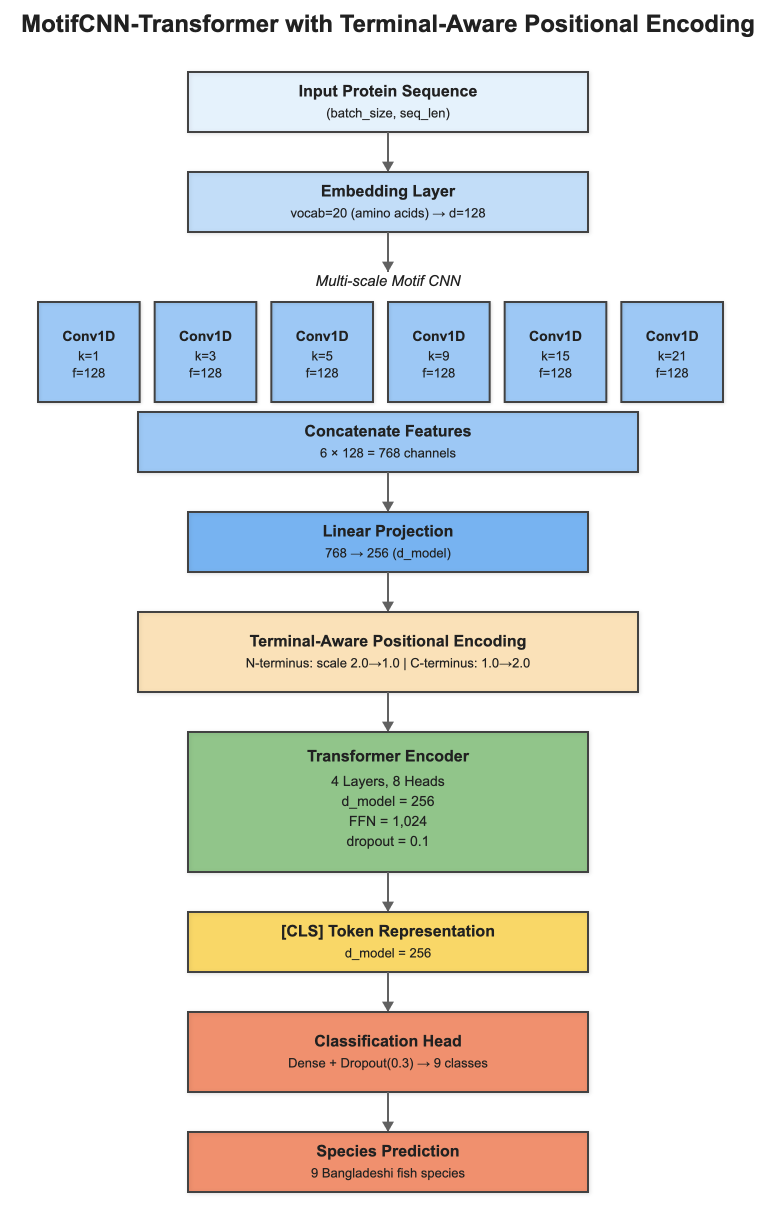}
    \caption{Architecture of the proposed MotifCNN-Transformer+TA-PE
    model showing the integration of multiscale convolutional feature extraction, Terminal-Aware Positional-Encoding, and Transformer based sequential modeling for protein based fish species identification.}
    \label{fig:architecture}
\end{figure}

6) MotifCNN-Transformer with Terminal-Aware Positional-Encoding: Our novel hybrid architecture illustrated in Fig.~\ref{fig:architecture} integrates three synergistic components: i) a multiscale MotifCNN explained in previously in point 2, frontend with six parallel 1D convolutions(kernel sizes: 1, 3, 5, 9, 15, 21, filters: 128 for each kernel), 
ii) a compact 4 layer Transformer encoder(embedding dimension=256, 8 attention heads, feedforward dimension=1024, dropout=0.1), 
and iii) Terminal-Aware Positional-Encoding explained previously that applies biologically motivated scaling to emphasize functionally critical N terminus and C terminus regions. The multiscale CNN features are concatenated(768 channels) and projected to 256 dimensions before entering the Transformer that enables hierarchical feature learning from local motifs to global patterns. A classification head with dropout(0.3) processes the CLS token representation for final prediction.

7) Finetuned ProtBERT: We finetune ProtBERT~\cite{r2} which is a 420 million parameter model (30 layers, embedding dimension=1024, 16 attention heads, feedforward dimension=4096) pretrained on 217 million protein sequences. Only this architecture marks the advent of pretraining. A classification layer was added for 9 classes.

All models were evaluated using weighted training. Inverse class frequency weighting was applied to address the inherent class imbalance in the dataset. This weighting scheme ensures proportional contribution of underrepresented species during optimization to prevent model bias towards more abundant classes. Accuracy, macro-$F_1$, micro-$F_1$, per-class performance measures and the same Cross Entropy Loss function served as consistence metrics across all models. Deep learning models were implemented in PyTorch 3.0 for 100 epochs with early stopping (patience=10) based on validation performance. They were trained under a single GPU(except ProtBERT) using the same AdamW optimizer with learning rate $1 \times 10^{-5}$ for ProtBERT, $1 \times 10^{-4}$ for others and weight decay of 0.01 . Batch size varies across different architectures: 32 for CNN-based models, 16 for standard Transformers, 64 for MotifCNN-Transformer, and 4 for ProtBERT. ProtBERT requires dual 16GB GPUs and high end CPU. Hyperparameters were selected through systematic grid search on the validation split. We used random seed 42 for reproducibility.

\section{Results and Analysis}

Table~\ref{tab:leaderboard} presents our comprehensive benchmark results, revealing distinct performance characteristics across model families along with their inference time. ProtBERT achieves highest 83.04\% accuracy but requires 5$\times$ more inference time than our MotifCNN-Transformer+TA-PE which achieves 79.80\% accuracy. Despite ProtBERT's 3.24\% higher accuracy, McNemar's test reveals that the difference between ProtBERT and our MotifCNN-Transformer+TA-PE is not statistically significant ($p=0.1120$).

\begin{table}[ht]
\centering
\caption{Comprehensive benchmark results on test set. Accuracy, macro-$F_1$ and inference time(s)}
\label{tab:leaderboard}
\begin{tabularx}{\columnwidth}{Xccc}
\toprule
Model & Accuracy & macro-$F_1$ & Inference \\
\midrule
N-gram + LightGBM & 0.4411 & 0.438 & 2.8 \\
MotifCNN & 0.637 & 0.6302 & 1.0 \\
CNN + BiLSTM & 0.6178 & 0.610 & 3.0 \\
Transformer (Standard PE) & 0.7105 & 0.690 & 3.5 \\
Transformer (TA-PE) & 0.752 & 0.7498 & 3.5 \\
\textbf{MotifCNN-Transformer+TA-PE} & \textbf{0.7980} & \textbf{0.800} & \textbf{4.2} \\
\textbf{ProtBERT} & \textbf{0.8304} & \textbf{0.830} & \textbf{21} \\
\bottomrule
\end{tabularx}
\end{table}

The results in table~\ref{tab:classwise_f1} illustrates class-wise F1-score comparisons for ProtBERT and MotifCNN-Transformer, highlighting species-specific performance patterns. Although ProtBERT achieves superior overall performance, our MotifCNN-Transformer+TA-PE outperforms it in species specific classification across six of nine species (baliya, chital, ilish, pangas, rupchanda, and shapla pata). The largest performance gaps favoring ProtBERT occur for near phylogenetic cyprinid species mrigal (+0.1877) and rohi (+0.2273). This gap suggests that extensive pretraining provides critical advantages for discriminating phylogenetically proximate species.

\begin{table}[ht]
\centering
\caption{Class-wise F1-score comparison of ProtBERT and MotifCNN-Transformer models.}
\label{tab:classwise_f1}
\begin{tabularx}{\columnwidth}{l c c X}
\toprule
Class & ProtBERT & MotifCNN-Transformer+TA-PE & Difference \\
\midrule
baliya & 0.8485 & 0.8615 & -0.0131 \\
chital & 0.8710 & 0.8760 & -0.0051 \\
ilish & 0.8267 & 0.8955 & -0.0689 \\
mrigal & 0.8271 & 0.6393 & +0.1877 \\
pangas & 0.8372 & 0.8916 & -0.0544 \\
rohi & 0.7500 & 0.5227 & +0.2273 \\
rupchanda & 0.7632 & 0.8148 & -0.0517 \\
shapla pata & 0.8657 & 0.9538 & -0.0882 \\
sing & 0.7458 & 0.7241 & +0.0216 \\
\bottomrule
\end{tabularx}
\end{table}

\begin{figure}[ht!]
    \centering
    \includegraphics[width=\linewidth]{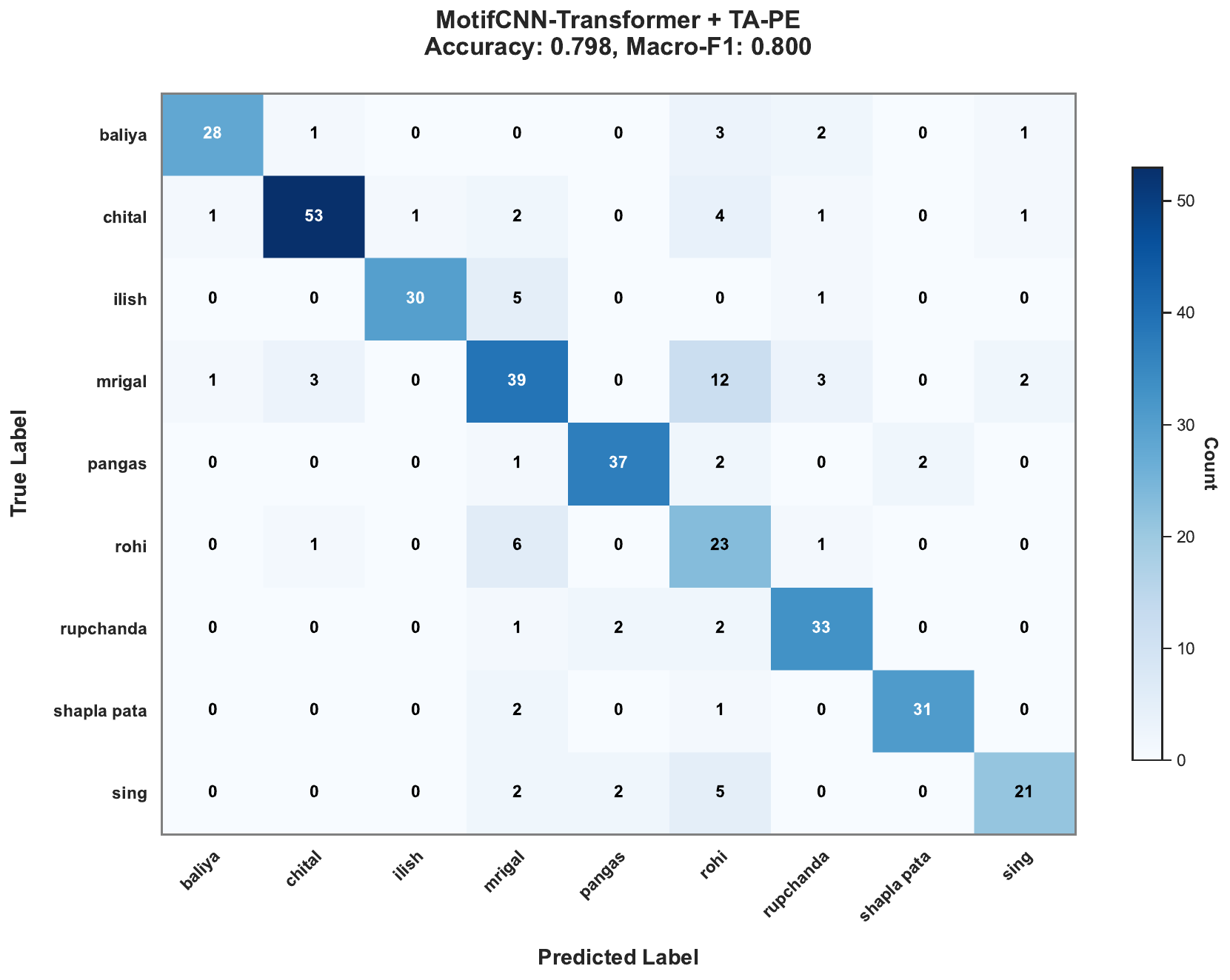}
    \caption{Confusion matrix for MotifCNN-Transformer revealing phylogenetically coherent discrimination patterns.}
    \label{fig:confusion_motifcnn}
\end{figure}

\begin{figure}[ht!]
    \centering
    \includegraphics[width=\linewidth]{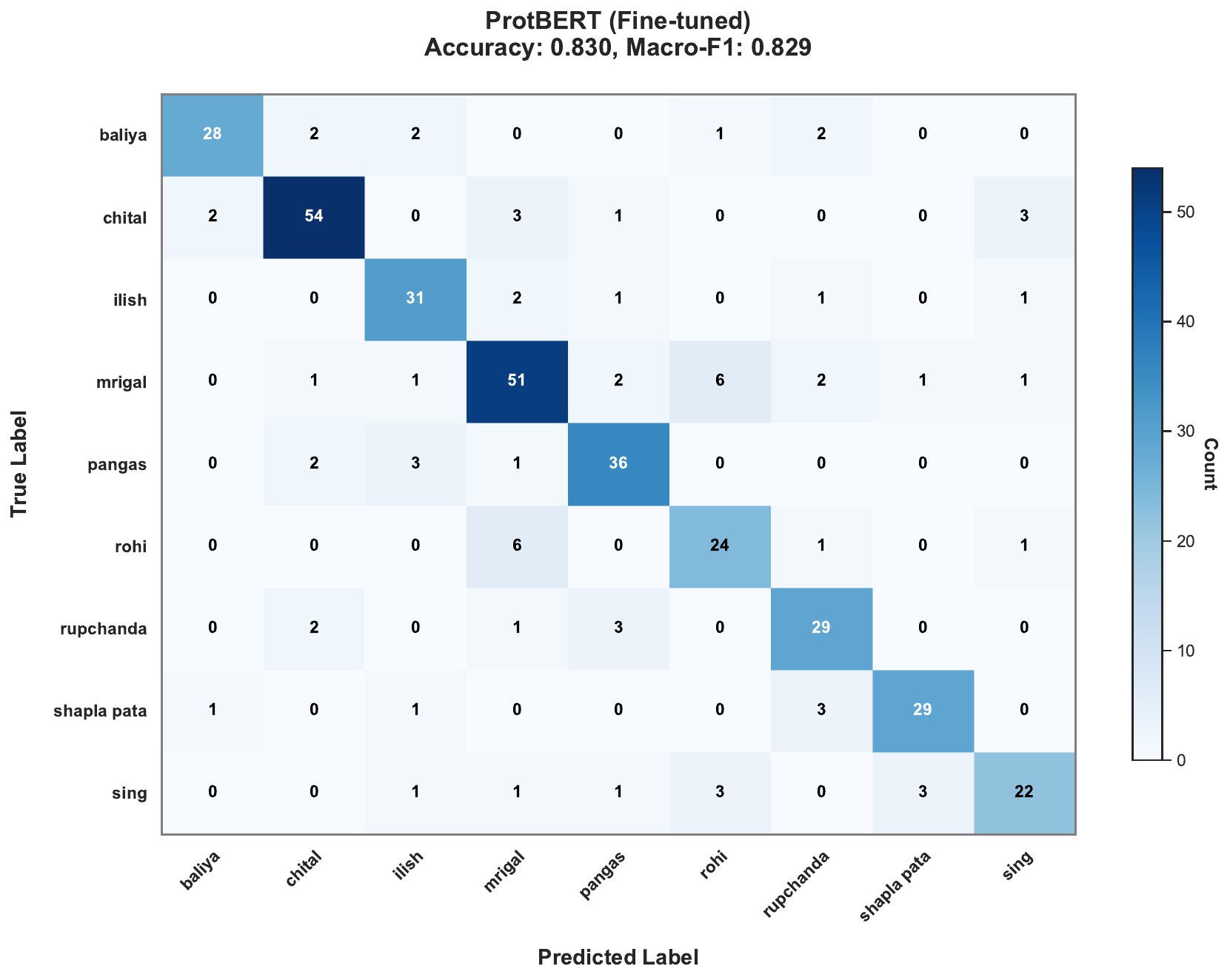}
    \caption{Confusion matrix for ProtBERT showing enhanced discrimination among closely related species while maintaining taxonomically coherent error patterns.}
    \label{fig:confusion_protbert}
\end{figure}

Our species level analysis reveals distinct performance patterns with profound biological implications. The confusion matrix in Fig.~\ref{fig:confusion_motifcnn} of MotifCNN-Transformer+TA-PE demonstrates taxonomically coherent discrimination capabilities. ProtBERT demonstrates improved discrimination capability for closely related species such as cyprinids in the confusion matrix of Fig.~\ref{fig:confusion_protbert}. Both models exhibit similar similar error patterns. Most of the misclassifications occur between near phylogenetic cyprinid species rohi(Labeo rohita) and mrigal(Cirrhinus mrigala). Both models achieves high classification accuracy(recall > 90\%) for distant phylogenetic species such as chital(Chitala chitala), pangas(Pangasius pangasius), ilish(Tenualosa ilisha) and shapla pata(Himantura uarnak).

Three key insights emerge from synthesizing these species specific patterns: 1) deep models consistently surpass classical baselines, 2) attention mechanisms outperform recurrent architectures for sequence modeling, and 3) biologically grounded inductive biases deliver incremental gains. Our hybrid MotifCNN-Transformer+TA-PE achieved the strongest non PLM result (79.80\%) while retaining substantial efficiency advantages compared to ProtBERT (83.04\%). Together with conservation analysis, these findings indicate that evolutionary constraints fundamentally bound separability for closely related phylogenetic species.

\section{Discussion}

Our benchmark reveals a clear performance hierarchy shaped fundamentally by sequence similarity constraints. Starting from the baseline N-gram + LightGBM (44.11\%), the CNN + BiLSTM architecture gains +17.67\% accuracy by modeling sequential dependencies, yet struggles with highly similar cyprinid species due to sequence homology exceeding 85\%. The MotifCNN surpasses it by +1.92\% through multi-scale local feature extraction. The standard Transformer achieves a substantial +7.35\% gain by capturing global dependencies that help discriminate moderately similar species. Incorporating biologically motivated Terminal-Aware Positional-Encoding further improves accuracy by +4.15\%, as terminal regions provide critical discriminative signals where central domains exhibit high conservation.

The proposed MotifCNN-Transformer+TA-PE architecture achieves 79.80\% accuracy, only 3.24\% below ProtBERT's 83.04\% despite being 42$\times$ smaller, supporting 16$\times$ larger batch size and requiring no GPU acceleration with 5$\times$ speed. Crucially, this difference is statistically insignificant ($p=0.1120$) and our hybrid model outperforms ProtBERT in six species with lower sequence similarity(baliya, chital, ilish, pangas, rupchanda, shapla pata). However, ProtBERT's pretraining advantage yields +22.73\% and +18.77\% F1 gains for highly homologous cyprinids rohi and mrigal respectively. This gain demonstrates that sequence similarity fundamentally bounds performance without excessive pretraining. Thiese results establish that our architecture achieves optimal performance within biological constraints while maintaining practical deployability in resource constrained environments where GPU infrastructure is intractable.

Our efficient architecture for real world fisheries management maximizes performance within the boundaries of evolutionary constrains imposing theoretical limits on discriminating species with >85\% sequence similarity. The 90-95\% accuracy achieved for phylogenetically distant species demonstrates that sequence based methods remain highly effective for most authentication scenarios. Future work should integrate multimodal data(spectroscopic, chemical) to overcome sequence similarity limitations for closely related species. Also coverage to Bangladesh's 260+ native fish species should be expanded. The computational efficiency and CPU only inference capability of our model make it uniquely suitable for field deployment in fishing communities.

\section{Conclusion}

This work establishes the foundation for protein-based identification of Bangladeshi fish species. We addressed a critical gap in regional biological datasets and presented the first curated protein sequence dataset for nine native Bangladeshi fish species. We also established a performance benchmark through comprehensive evaluation of six modeling paradigms.

Our novel MotifCNN-Transformer architecture with Terminal-Aware Positional-Encoding achieves 79.80\% accuracy with batch sizes up to 64 and enables GPU free inference requiring modest computational resources that provides significant computational advantages over alternatives. These efficiency gains are crucial compared to ProtBERT, which achieves state of the art 83.04\% accuracy but requires dual 16GB GPUs, high-end CPUs, batch size support limited to 4, and 80+ minutes of inference times. These requirements exhibit ProtBERT's impracticality for deployment in resource constrained environments such as rural Bangladesh where species identification is most critical. 

Our work enables automated species identification systems by providing practically deployable computationally efficient models along with foundational data resources. These resources could transform fisheries management, combat food fraud, and support biodiversity conservation in one of the world's most fish dependent nations. Our contributions establish a foundation for addressing these challenges and provides a solution for the global fish stock which is threaten by climate change and overfishing.

\bibliographystyle{IEEEtran}
\bibliography{references}

\begin{thebibliography}{10}
\providecommand{\url}[1]{#1}
\csname url@samestyle\endcsname
\providecommand{\newblock}{\relax}
\providecommand{\bibinfo}[2]{#2}
\providecommand{\BIBentrySTDinterwordspacing}{\spaceskip=0pt\relax}
\providecommand{\BIBentryALTinterwordstretchfactor}{4}
\providecommand{\BIBentryALTinterwordspacing}{\spaceskip=\fontdimen2\font plus
\BIBentryALTinterwordstretchfactor\fontdimen3\font minus \fontdimen4\font\relax}
\providecommand{\BIBforeignlanguage}[2]{{%
\expandafter\ifx\csname l@#1\endcsname\relax
\typeout{** WARNING: IEEEtran.bst: No hyphenation pattern has been}%
\typeout{** loaded for the language `#1'. Using the pattern for}%
\typeout{** the default language instead.}%
\else
\language=\csname l@#1\endcsname
\fi
#2}}
\providecommand{\BIBdecl}{\relax}
\BIBdecl

\bibitem{r30}
\BIBentryALTinterwordspacing
A.~R. Sunny, M.~H. Mithun, S.~H. Prodhan, M.~Ashrafuzzaman, S.~M.~A. Rahman, M.~M. Billah, M.~Hussain, K.~J. Ahmed, S.~A. Sazzad, M.~T. Alam, A.~Rashid, and M.~M. Hossain, ``Fisheries in the context of attaining sustainable development goals (sdgs) in bangladesh: Covid-19 impacts and future prospects,'' \emph{Sustainability}, vol.~13, no.~17, 2021. [Online]. Available: \url{https://www.mdpi.com/2071-1050/13/17/9912}
\BIBentrySTDinterwordspacing

\bibitem{ZHAO2024}
\BIBentryALTinterwordspacing
S.~Zhao, H.~Zhang, Z.~Zhao, Y.~Zhang, J.~Yu, Y.~Tang, and C.~Zhou, ``Integrated dna barcoding methods to identify species in the processed fish products from chinese market,'' \emph{Food Research International}, vol. 182, p. 114140, 2024. [Online]. Available: \url{https://www.sciencedirect.com/science/article/pii/S0963996924002102}
\BIBentrySTDinterwordspacing

\bibitem{Meledina2025}
\BIBentryALTinterwordspacing
A.~Meledina, D.~Straka, F.~Soucek, T.~A. Smirnova, and S.~Kuckova, ``Rapid determination of fish species of raw and heat-treated fish meat using proteomic species-specific markers,'' \emph{Food Technology and Biotechnology}, vol.~63, no.~3, pp. 287--297, Jul--Sep 2025, epub 2025 Aug 31. PMID: 41000212; PMCID: PMC12413489. [Online]. Available: \url{https://pubmed.ncbi.nlm.nih.gov/41000212/}
\BIBentrySTDinterwordspacing

\bibitem{CHIEN2022}
\BIBentryALTinterwordspacing
H.-J. Chien, Y.-H. Huang, Y.-F. Zheng, W.-C. Wang, C.-Y. Kuo, G.-J. Wei, and C.-C. Lai, ``Proteomics for species authentication of cod and corresponding fishery products,'' \emph{Food Chemistry}, vol. 374, p. 131631, 2022. [Online]. Available: \url{https://www.sciencedirect.com/science/article/pii/S0308814621026376}
\BIBentrySTDinterwordspacing

\bibitem{Irschick2017}
\BIBentryALTinterwordspacing
D.~J. Irschick, A.~Fu, G.~Lauder, C.~Wilga, C.-Y. Kuo, and N.~Hammerschlag, ``A comparative morphological analysis of body and fin shape for eight shark species,'' \emph{Biological Journal of the Linnean Society}, vol. 122, no.~3, pp. 589--604, 08 2017. [Online]. Available: \url{https://doi.org/10.1093/biolinnean/blx088}
\BIBentrySTDinterwordspacing

\bibitem{Adhikary2025}
\BIBentryALTinterwordspacing
S.~Adhikary, S.~Banerjee, R.~Singh, and A.~D. Dwivedi, ``Fish species identification on low resolution—a study with enhanced super-resolution generative adversarial network (esrgan), yolo and vgg-16,'' \emph{PeerJ Computer Science}, vol.~11, p. e2860, 2025. [Online]. Available: \url{https://pubmed.ncbi.nlm.nih.gov/40567807/}
\BIBentrySTDinterwordspacing

\bibitem{Xing2018}
\BIBentryALTinterwordspacing
X.~Bingpeng, L.~Heshan, Z.~Zhilan, W.~Chunguang, W.~Yanguo, and W.~Jianjun, ``Dna barcoding for identification of fish species in the taiwan strait,'' \emph{PLOS ONE}, vol.~13, no.~6, pp. 1--13, 06 2018. [Online]. Available: \url{https://doi.org/10.1371/journal.pone.0198109}
\BIBentrySTDinterwordspacing

\bibitem{Quek2024}
\BIBentryALTinterwordspacing
Z.~B.~R. Quek, Z.~T. Yip, S.~S. Jain, H.~X.~V. Wong, Z.~Tan, A.~R. Joseph, and D.~Huang, ``Dna barcodes are ineffective for species identification of acropora corals from the aquarium trade,'' \emph{Biodiversity Data Journal}, vol.~12, p. e125914, 2024. [Online]. Available: \url{https://doi.org/10.3897/BDJ.12.e125914}
\BIBentrySTDinterwordspacing

\bibitem{chakraborty2025}
\BIBentryALTinterwordspacing
P.~Chakraborty and A.~Bhargava, ``Xai-driven deep learning for protein sequence functional group classification,'' 2025. [Online]. Available: \url{https://arxiv.org/abs/2511.13791}
\BIBentrySTDinterwordspacing

\bibitem{Ahmed2024}
N.~Y. Ahmed, W.~A. Alsanousi, E.~M. Hamid, M.~K. Elbashir, K.~M. Al-Aidarous, M.~Mohammed, and M.~E.~M. Musa, ``An efficient deep learning approach for dna-binding proteins classification from primary sequences,'' \emph{International Journal of Computational Intelligence Systems}, vol.~17, no.~1, p.~88, Apr. 2024.

\bibitem{Hein2025}
\BIBentryALTinterwordspacing
Z.~M. Hein, D.~Guruparan, B.~Okunsai, C.~M.~N. Che Mohd~Nassir, M.~D.~C. Ramli, and S.~Kumar, ``Ai and machine learning in biology: From genes to proteins,'' \emph{Biology}, vol.~14, no.~10, 2025. [Online]. Available: \url{https://www.mdpi.com/2079-7737/14/10/1453}
\BIBentrySTDinterwordspacing

\bibitem{Bari2025}
\BIBentryALTinterwordspacing
P.~Bari, G.~Bedi, K.~Joshi, and A.~Jawale, ``Why transformers outperform lstms: A comparative study on sarcasm detection,'' \emph{Journal on Artificial Intelligence}, vol.~7, no.~1, pp. 499--508, 2025. [Online]. Available: \url{http://www.techscience.com/jai/v7n1/64530}
\BIBentrySTDinterwordspacing

\bibitem{Lin2023}
\BIBentryALTinterwordspacing
Z.~Lin, H.~Akin, R.~Rao, B.~Hie, Z.~Zhu, W.~Lu, N.~Smetanin, R.~Verkuil, O.~Kabeli, Y.~Shmueli, A.~dos Santos~Costa, M.~Fazel-Zarandi, T.~Sercu, S.~Candido, and A.~Rives, ``Evolutionary-scale prediction of atomic-level protein structure with a language model,'' \emph{Science}, vol. 379, no. 6637, pp. 1123--1130, 2023. [Online]. Available: \url{https://www.science.org/doi/abs/10.1126/science.ade2574}
\BIBentrySTDinterwordspacing

\bibitem{Mahala2025}
\BIBentryALTinterwordspacing
A.~Mahala, A.~Ranjan, R.~Priyadarshini, R.~Vikram, and P.~Dansena, ``A fast (cnn + mcws-transformer) based architecture for protein function prediction,'' \emph{Statistical Applications in Genetics and Molecular Biology}, vol.~24, no.~1, Jul. 2025, pMID: 40586353. [Online]. Available: \url{https://pubmed.ncbi.nlm.nih.gov/40586353/}
\BIBentrySTDinterwordspacing

\bibitem{sankar2019}
\BIBentryALTinterwordspacing
A.~Sankar, X.~Zhang, and K.~C.-C. Chang, ``Motif-based convolutional neural network on graphs,'' 2019. [Online]. Available: \url{https://arxiv.org/abs/1711.05697}
\BIBentrySTDinterwordspacing

\bibitem{Choi2023}
\BIBentryALTinterwordspacing
S.~R. Choi and M.~Lee, ``Transformer architecture and attention mechanisms in genome data analysis: A comprehensive review,'' \emph{Biology}, vol.~12, no.~7, 2023. [Online]. Available: \url{https://www.mdpi.com/2079-7737/12/7/1033}
\BIBentrySTDinterwordspacing

\bibitem{r29}
K.~Habib and M.~ISLAM, ``An updated checklist of marine fishes of bangladesh,'' \emph{Bangladesh Journal of Fisheries}, vol.~32, pp. 357--367, 01 2021.

\bibitem{r27}
M.~S. Ahmed, M.~Chowdhury, and L.~Nahar, ``Molecular characterization of small indigenous fish species (sis) of bangladesh through dna barcodes,'' \emph{Gene}, vol. 684, 10 2018.

\bibitem{r28}
P.~Das, M.~Kawsar, P.~Biswas~Paul, A.~A. Hridoy, M.~Hossain, and S.~Niloy, ``Bd-freshwater-fish: An image dataset from bangladesh for ai-powered automatic fish species classification and detection toward smart aquaculture,'' \emph{Data in Brief}, vol.~57, p. 111132, 11 2024.

\bibitem{UniProt2024}
\BIBentryALTinterwordspacing
T.~U. Consortium, ``Uniprot: the universal protein knowledgebase in 2025,'' \emph{Nucleic Acids Research}, vol.~53, no.~D1, pp. D609--D617, 11 2024. [Online]. Available: \url{https://doi.org/10.1093/nar/gkae1010}
\BIBentrySTDinterwordspacing

\bibitem{Sayers2025}
\BIBentryALTinterwordspacing
E.~W. Sayers \emph{et~al.}, ``Database resources of the national center for biotechnology information in 2025,'' \emph{Nucleic Acids Research}, vol.~53, no.~D1, pp. D20--D29, 2025. [Online]. Available: \url{https://pubmed.ncbi.nlm.nih.gov/39526373/}
\BIBentrySTDinterwordspacing

\bibitem{MARSHALL2023}
\BIBentryALTinterwordspacing
A.~C. Marshall, J.~Cummins, S.~Kobelke, T.~Zhu, J.~Widagdo, V.~Anggono, A.~Hyman, A.~H. Fox, C.~S. Bond, and M.~Lee, ``Different low-complexity regions of sfpq play distinct roles in the formation of biomolecular condensates,'' \emph{Journal of Molecular Biology}, vol. 435, no.~24, p. 168364, 2023. [Online]. Available: \url{https://www.sciencedirect.com/science/article/pii/S0022283623004758}
\BIBentrySTDinterwordspacing

\bibitem{Berrow2021}
N.~Berrow, A.~de~Marco, M.~Lebendiker, M.~Garcia-Alai, S.~H. Knauer, B.~Lopez-Mendez, A.~Matagne, A.~Parret, K.~Remans, S.~Uebel, and B.~Raynal, ``Quality control of purified proteins to improve data quality and reproducibility: results from a large-scale survey,'' \emph{European Biophysics Journal}, vol.~50, no.~3, pp. 453--460, May 2021.

\bibitem{Paul2008}
S.~Paul, S.~K. Bag, S.~Das, E.~T. Harvill, and C.~Dutta, ``Molecular signature of hypersaline adaptation: insights from genome and proteome composition of halophilic prokaryotes,'' \emph{Genome Biology}, vol.~9, no.~4, p. R70, Apr. 2008.

\bibitem{Oren2013}
\BIBentryALTinterwordspacing
A.~Oren, ``Life at high salt concentrations, intracellular kcl concentrations, and acidic proteomes,'' \emph{Frontiers in Microbiology}, vol. Volume 4 - 2013, 2013. [Online]. Available: \url{https://www.frontiersin.org/journals/microbiology/articles/10.3389/fmicb.2013.00315}
\BIBentrySTDinterwordspacing

\bibitem{DeepLoc2017}
J.~Armenteros, C.~Sønderby, S.~Sønderby, H.~Nielsen, and O.~Winther, ``Deeploc: prediction of protein subcellular localization using deep learning,'' \emph{Bioinformatics (Oxford, England)}, vol.~33, 07 2017.

\bibitem{r2}
\BIBentryALTinterwordspacing
N.~Brandes, D.~Ofer, Y.~Peleg, N.~Rappoport, and M.~Linial, ``Proteinbert: a universal deep-learning model of protein sequence and function,'' \emph{Bioinformatics}, vol.~38, no.~8, pp. 2102--2110, 02 2022. [Online]. Available: \url{https://doi.org/10.1093/bioinformatics/btac020}
\BIBentrySTDinterwordspacing

\end{thebibliography}
\end{document}